\documentclass[aps,preprint]{revtex4}%
\usepackage{amsmath}
\usepackage{graphicx}
\usepackage{amsfonts}
\usepackage{amssymb}%
\begin{document}
\title{Bilinear and quadratic Hamiltonians in two-mode cavity quantum electrodynamics }
\author{F. O. Prado$^{1}$, N. G. de Almeida$^{2}$, M. H. Y. Moussa$^{1}$ and C. J.
Villas-B\^{o}as$^{1}$ }
\affiliation{$^{1}$Universidade Federal de S\~{a}o Carlos, S\~{a}o Carlos, Brazil }
\affiliation{$^{2}$Universidade Cat\'{o}lica de Goi\'{a}s, Goi\^{a}nia, Brazil}

\begin{abstract}
In this work we show how to engineer bilinear and quadratic Hamiltonians in
cavity quantum electrodynamics (QED) through the interaction of a single
driven two-level atom with cavity modes. The validity of the engineered
Hamiltonians is numerically analyzed even considering the effects of \ both
dissipative mechanisms, the cavity field and the atom. The present scheme can
be used, in both optical and microwave regimes, for quantum state preparation,
the implementation of quantum logical operations, and fundamental tests of
quantum theory.

\end{abstract}
\maketitle

\address{$^{1}$Departamento de F\'{\i}sica, Universidade Federal de S\~{a}o\\
Carlos, P.O. Box 676, S\~{a}o Carlos\textit{, 13565-905, }S\~{a}oPaulo,\textit{\ }Brazil\\
$^{2}$Departamento de Matem\'{a}tica e F\'{\i}sica, Universidade Cat\'{o}lica
de Goi\'{a}s, P.O. Box 86, Goi\^{a}nia,74605-010, Goi\'{a}s, Brazil}

\section{Introduction}

\bigskip

Frequency-conversion mechanisms, such as optical parametric and four-wave
mixing processes, have acted as a basic resource in the investigation of
fundamental quantum phenomena over the last few decades. Largely employed to
produce squeezed and polarization-entangled photon states to test
sub-Poissonian statistics \cite{Stoler} and Bell's inequalities \cite{Kwiat}%
,\ such processes have deepened our understanding of radiation \cite{Stoler}
and its interaction with matter \cite{Milburn}. Apart from applications in
fundamental physics, it has been conjectured that frequency conversions can
improve the signal-to-noise ratio in optical communication \cite{SN} and be
used to measure gravitational waves through squeezed fields \cite{Caves}.
Recently, they have also been required within quantum information theory for
the implementation of a nondeterministic controlled-NOT operation
\cite{Pittman}.

Against this backdrop of the general usefulness of frequency conversions in
the running-wave domain, several recent studies have been devoted to\ mapping
these mechanisms into two-mode cavity quantum electrodynamics (QED)
\cite{Celso1,Roberto,Celso2,Zagury}. Parametric up- and down-conversions (PUC
and PDC) were accomplished through the dispersive interactions of a single
three-level atom simultaneously with a classical driving field and a two-mode
cavity. The PDC (PUC) process follows from the ladder (lambda) configuration
of atomic levels, in which the ground $\left\vert g\right\rangle $ and excited
$\left\vert e\right\rangle $ states are coupled through an auxiliary
intermediate (more-excited) level $\left\vert i\right\rangle $. The cavity
modes $\omega_{a}$ and $\omega_{b}$ are tuned to the vicinity of the
dipole-allowed transitions $\left\vert g\right\rangle $ $\leftrightarrow$
$\left\vert i\right\rangle $ and $\left\vert e\right\rangle $ $\leftrightarrow
$ $\left\vert i\right\rangle $. The desired interaction between the modes
$\omega_{a}$ and $\omega_{b}$, $\hbar\left(  \xi ab+\xi^{\ast}a^{\dagger
}b^{\dagger}\right)  $ for PDC or $\hbar\left(  \xi ab^{\dagger}+\xi^{\ast
}a^{\dagger}b\right)  $ for PUC, is accomplished by driving the
dipole-forbidden atomic transition $\left\vert g\right\rangle $
$\leftrightarrow$ $\left\vert e\right\rangle $ out of resonance with a
classical field. For the degenerate PDC process, where $\omega_{a}=\omega_{b}%
$, the well-known interaction $\hbar\left[  \xi\left(  a\right)  ^{2}%
+\xi^{\ast}\left(  a^{\dagger}\right)  ^{2}\right]  $ was first achieved in
\cite{Celso1}, and this may be used to squeeze an arbitrary state previously
prepared in the cavity; i.e., to perform the squeezing operation $S\left\vert
\Psi\right\rangle $ in cavity QED ($S$ being the squeeze operator). These
achievements enhance prospects of quantum information manipulation and of
fundamental tests of quantum theory in cavity QED. In fact, the engineered
bilinear Hamiltonians can be used to generate one-mode mesoscopic squeezed
superpositions, two-mode entanglements, and two-mode squeezed vacuum states
(such the original EPR state).

Motivated by these accomplishments \cite{Celso1,Roberto,Celso2,Zagury}, and
simultaneously attempting to generalize and simplify these protocols, in the
present letter we consider only a two-level Rydberg atom in order to generate,
in two-mode cavity QED, bilinear and quadratic Hamiltonians similar to those
describing PUC and PDC. We also demonstrate how to generate, in one-mode
cavity QED, the anti-Jaynes-Cummings (AJC) Hamiltonian \cite{Solano} and a
mixture of the Jaynes-Cummings (JC) and the AJC Hamiltonians. We stress that
all the previous schemes presented in literature to generate nonlinear
Hamiltonians consider the interaction between a cavity mode and an atom with
at least three atomic levels. The present protocol, overcomes the difficulty
of driving the dipole-forbidden atomic transition $\left\vert g\right\rangle $
$\leftrightarrow$ $\left\vert e\right\rangle $, which sometimes requires a
significant strength, as in the strong amplification regime defined in Ref.
\cite{Celso2}. More importantly, with a two-level atom, long-lived circular
Rydberg states might be employable reducing the noise coming from the finite
lifetimes of the atomic levels. (We stress that, to achieve PUC and PDC with a
three-level atomic configuration, at least one level cannot be a long-lived
circular Rydberg state.) Our results are derived from variations of the
Hamiltonian $H=H_{0}+V(t)$ (with $\hbar=1$), with
\begin{subequations}
\label{1}%
\begin{align}
H_{0}  &  =\omega_{a}a^{\dagger}a+\omega_{b}b^{\dagger}b+\omega_{0}\left(
\sigma_{ee}-\sigma_{gg}\right)  /2,\label{1a}\\
V(t)  &  =\left[  \lambda_{a}a\sigma_{eg}+\lambda_{b}b\sigma_{eg}+\left(
\Omega_{1}\operatorname*{e}\nolimits^{-i\omega_{1}t}+\Omega_{2}%
\operatorname*{e}\nolimits^{-i\omega_{2}t}\right)  \sigma_{eg}+\mathrm{h{.c.}%
}\right]  {,} \label{1b}%
\end{align}
where the atomic ground ($g$) and excited ($e$) states, with transition
frequency $\omega_{0}$, are coupled non-resonantly through the cavity modes of
frequencies $\omega_{a}$ and $\omega_{b}$, with coupling constants
$\lambda_{a}$ and $\lambda_{b}$, and detunings $\delta_{a}=\omega_{0}%
-\omega_{a}$ and $\delta_{b}$ $=\omega_{0}-\omega_{b}$. The dipole-allowed
transition $\left\vert g\right\rangle $ $\leftrightarrow$ $\left\vert
e\right\rangle $ is also excited by two driving classical fields of
frequencies $\omega_{1}$ and $\omega_{2}$, with coupling constants $\Omega
_{1}=\left\vert \Omega_{1}\right\vert \operatorname*{e}^{i\varphi_{1}}$ and
$\Omega_{2}=\left\vert \Omega_{2}\right\vert \operatorname*{e}^{i\varphi_{2}}%
$, and detunings $\delta_{1}=\omega_{0}-\omega_{1}$ and $\delta_{2}=\omega
_{0}-\omega_{2}$. In the interaction picture, the transformed Hamiltonian is
given by
\end{subequations}
\begin{equation}
V_{I}(t)=\lambda_{a}\operatorname*{e}\nolimits^{i\delta_{a}t}a\sigma
_{eg}+\lambda_{b}\operatorname*{e}\nolimits^{i\delta_{b}t}b\sigma_{eg}+\left(
\Omega_{1}\operatorname*{e}\nolimits^{i\delta_{1}t}+\Omega_{2}%
\operatorname*{e}\nolimits^{i\delta_{2}t}\right)  \sigma_{eg}+\mathrm{h{.c.}%
}\text{.} \label{1c}%
\end{equation}

\section{\textit{Bilinear and quadratic Hamiltonians}}

A single classical amplification field ($\omega_{1}$) is required to
accomplish these interactions. After writing the Hamiltonian $V_{I}(t)$ in
this laser framework and defining a new basis for the atomic states $\left\{
\left|  \pm\right\rangle \simeq\left(  \operatorname*{e}\nolimits^{i\varphi
_{1}}\left|  e\right\rangle \pm\left|  g\right\rangle \right)  /\sqrt
{2}\right\}  $\cite{Solano}, under the assumption that $\delta_{1}\ll$
$\left|  \Omega_{1}\right|  $, $\left|  \delta_{a}\right|  $, $\left|
\delta_{b}\right|  $, we proceed to the transformation $\mathcal{U}$
$=\exp\left[  -i\left|  \Omega_{1}\right|  \left(  \sigma_{++}-\sigma
_{--}\right)  t\right]  $, which prepares the Hamiltonian
\begin{equation}
\mathcal{V}(t)=\left(  \widetilde{\lambda}_{a}\operatorname*{e}%
\nolimits^{i\left(  \delta_{a}-\delta_{1}\right)  t}a+\widetilde{\lambda}%
_{b}\operatorname*{e}\nolimits^{i\left(  \delta_{b}-\delta_{1}\right)
t}b\right)  \left(  \sigma_{++}-\sigma_{--}-\operatorname*{e}%
\nolimits^{2i\Omega_{1}t}\sigma_{+-}+\operatorname*{e}\nolimits^{-2i\Omega
_{1}t}\sigma_{-+}\right)  /2+\mathrm{h{.c.}}\text{,} \label{1d}%
\end{equation}
with $\widetilde{\lambda}_{\alpha}=\lambda_{\alpha}\operatorname*{e}%
\nolimits^{-i\varphi_{1}}$ ($\alpha=a,b$), for a subsequent perturbation
approximation. We finally obtain, up to the second order, the effective
Hamiltonian \cite{James}
\begin{equation}
\mathcal{H}=-i\mathcal{V}(t)\int\mathcal{V}(t^{\prime})dt^{\prime}\text{,}
\label{1e}%
\end{equation}
which will be analysed under three regimes of the classical amplification
field: the weak ($\left|  \delta_{a}\right|  \sim\left|  \delta_{b}\right|
\gg$ $\left|  \Omega_{1}\right|  \gtrsim\left|  \widetilde{\lambda}%
_{a}\right|  \sim\left|  \widetilde{\lambda}_{b}\right|  $), the intermediate
($\left|  \Omega_{1}\right|  \sim\left|  \delta_{a}\right|  \sim\left|
\delta_{b}\right|  \gg\left|  \widetilde{\lambda}_{a}\right|  \sim\left|
\widetilde{\lambda}_{b}\right|  $) and the strong ($\left|  \Omega_{1}\right|
\gg\left|  \delta_{a}\right|  \sim\left|  \delta_{b}\right|  \gg\left|
\widetilde{\lambda}_{a}\right|  \sim\left|  \widetilde{\lambda}_{b}\right|  $)
amplification regimes.

It is important to note from Eq. (\ref{1d}) that once the relation $\left|
2\Omega_{1}\pm\left(  \delta_{\alpha}-\delta_{1}\right)  \right|  \gg\left|
\widetilde{\lambda}_{\alpha}\right|  $ is satisfied for all the amplification
regimes, there will be no transition between the atomic dressed states
$\left|  +\right\rangle $ and $\left|  -\right\rangle $ (even that there will
be transitions between the bare states $\left|  g\right\rangle $ and $\left|
e\right\rangle $).

\subsection{\textit{The Hamiltonian }$\left(  ab+\mathrm{h.c.}\right)  $}

From the above analysis, this interaction is achieved considering the energy
diagram pictorially sketched in Fig. 1(a), where $\delta_{a}=-$ $\delta
_{b}=\delta>0$. From Eq. (\ref{1e}), we obtain the effective Hamiltonian
$\mathcal{H}(t)=\mathcal{H}_{0}\mathcal{+H}_{int\text{ }}(t)$, with%
\begin{subequations}
\begin{align}
\mathcal{H}_{0}  &  =\frac{\left\vert \Omega_{1}\right\vert }{4\left\vert
\Omega_{1}\right\vert ^{2}-\delta^{2}}\left(  \left\vert \widetilde{\lambda
}_{a}\right\vert ^{2}a^{\dagger}a+\left\vert \widetilde{\lambda}%
_{b}\right\vert ^{2}b^{\dagger}b\right)  \left(  \sigma_{++}-\sigma
_{--}\right) \nonumber\\
&  +\frac{2}{\delta}\sum_{\ell=+,-}\left(  \left\vert \widetilde{\lambda}%
_{a}\right\vert ^{2}\frac{\delta+\ell\left\vert \Omega_{1}\right\vert }%
{\delta+2\ell\left\vert \Omega_{1}\right\vert }-\left\vert \widetilde{\lambda
}_{b}\right\vert ^{2}\frac{\delta-\ell\left\vert \Omega_{1}\right\vert
}{\delta-2\ell\left\vert \Omega_{1}\right\vert }\right)  \sigma_{\ell\ell
}\label{1la}\\
\mathcal{H}_{int\text{ }}(t)  &  =\left(  \frac{\widetilde{\lambda}%
_{a}\widetilde{\lambda}_{b}\left\vert \Omega_{1}\right\vert }{\delta
^{2}-4\left\vert \Omega_{1}\right\vert ^{2}}\operatorname*{e}%
\nolimits^{-2i\delta_{1}t}ab+\mathrm{h.c}\right)  \left(  \sigma_{++}%
-\sigma_{--}\right)  \text{.} \label{1lb}%
\end{align}
Preparing the atomic state $\left\vert \pm\right\rangle $, we obtain through
the unitary transformation $\mathcal{H}_{\pm}=U^{\dagger}\mathcal{H}%
(t)U-\mathcal{H}_{0}$, with $U=\operatorname*{e}\nolimits^{-i\mathcal{H}_{0}%
t}$, the engineered interaction
\end{subequations}
\begin{equation}
\mathcal{H}_{\pm}=\left(  \Lambda_{\pm}ab+\mathrm{h.c.}\right)  \mathrm{{,}}
\label{2}%
\end{equation}
where the coupling parameters in the weak ($W$), intermediate ($I$) and strong
($S$) amplification regimes become $\Lambda_{\pm W}=\pm\widetilde{\lambda}%
_{a}\widetilde{\lambda}_{b}\left\vert \Omega_{1}\right\vert /\delta^{2}$,
$\Lambda_{\pm I}=\pm\widetilde{\lambda}_{a}\widetilde{\lambda}_{b}\left\vert
\Omega_{1}\right\vert /\left(  \delta^{2}-4\left\vert \Omega_{1}\right\vert
^{2}\right)  $, and $\Lambda_{\pm S}=\mp\widetilde{\lambda}_{a}\widetilde
{\lambda}_{b}/4\left\vert \Omega_{1}\right\vert $, after adjusting the
detuning $\delta_{1}$ such that $\delta_{1W}=\pm\left(  \left\vert
\widetilde{\lambda}_{a}\right\vert ^{2}+\left\vert \widetilde{\lambda}%
_{b}\right\vert ^{2}\right)  \left\vert \Omega_{1}\right\vert /\delta^{2}$,
$\delta_{1I}=\pm\left(  \left\vert \widetilde{\lambda}_{a}\right\vert
^{2}+\left\vert \widetilde{\lambda}_{b}\right\vert ^{2}\right)  \left\vert
\Omega_{1}\right\vert /\left(  4\left\vert \Omega_{1}\right\vert ^{2}%
-\delta^{2}\right)  $, and $\delta_{1S}=\pm\left(  \left\vert \widetilde
{\lambda}_{a}\right\vert ^{2}+\left\vert \widetilde{\lambda}_{b}\right\vert
^{2}\right)  /4\left\vert \Omega_{1}\right\vert $, respectively. (For the
intermediate amplification regime care must be take to avoid the equality
$\left\vert \Omega_{1}\right\vert =\left\vert \delta\right\vert $.) We note
that the strength of the coupling parameters obey the relation $\Lambda
_{I}\gtrsim\Lambda_{W}\gtrsim\Lambda_{S}$.

\subsection{\textit{The Hamiltonian }$\left(  ab^{\dagger}+\mathrm{h.c.}%
\right)  $}

The energy diagram leading to this interaction is sketched in Fig. 1(b), where
$\delta_{a}\sim$ $\delta_{b}$. The Effective Hamiltonian becomes%
\begin{subequations}
\begin{align}
\mathcal{H}_{0}  &  =\left(  \frac{\left\vert \Omega_{1}\right\vert \left\vert
\widetilde{\lambda}_{a}\right\vert ^{2}}{4\left\vert \Omega_{1}\right\vert
^{2}-\delta_{a}^{2}}a^{\dagger}a+\frac{\left\vert \Omega_{1}\right\vert
\left\vert \widetilde{\lambda}_{b}\right\vert ^{2}}{4\left\vert \Omega
_{1}\right\vert ^{2}-\delta_{b}^{2}}b^{\dagger}b\right)  \left(  \sigma
_{++}-\sigma_{--}\right) \nonumber\\
&  +\frac{1}{2}\sum_{\ell=+,-}\left(  \left\vert \widetilde{\lambda}%
_{a}\right\vert ^{2}\frac{\delta_{a}+\ell\left\vert \Omega_{1}\right\vert
}{\delta_{a}\left(  \delta_{a}+2\ell\left\vert \Omega_{1}\right\vert \right)
}+\left\vert \widetilde{\lambda}_{b}\right\vert ^{2}\frac{\delta_{b}%
+\ell\left\vert \Omega_{1}\right\vert }{\delta_{b}\left(  \delta_{b}%
+2\ell\left\vert \Omega_{1}\right\vert \right)  }\right)  \sigma_{\ell\ell
}\text{,}\label{2a}\\
\mathcal{H}_{int\text{ }}(t)  &  =\frac{1}{2}\widetilde{\lambda}_{a}%
\widetilde{\lambda}_{b}^{\ast}\operatorname*{e}\nolimits^{i\left(  \delta
_{a}-\delta_{b}\right)  t}ab^{\dagger}\sum_{\ell=+,-}\left(  \frac{\delta
_{b}+\ell\left\vert \Omega_{1}\right\vert }{\delta_{b}\left(  \delta_{b}%
+2\ell\left\vert \Omega_{1}\right\vert \right)  }-\frac{\delta_{a}%
-\ell\left\vert \Omega_{1}\right\vert }{\delta_{a}\left(  \delta_{a}%
-2\ell\left\vert \Omega_{1}\right\vert \right)  }\right)  \sigma_{\ell\ell
}+\mathrm{h.c}\text{..} \label{2b}%
\end{align}
Again, preparing the atomic state $\left\vert \pm\right\rangle $ we obtain,
through the same steps leading to the interaction $\left(  ab+\mathrm{h.c.}%
\right)  $, the effective Hamiltonian
\end{subequations}
\begin{equation}
\mathcal{H}_{\pm}=\left(  \Sigma_{\pm}ab^{\dagger}\operatorname*{e}%
\nolimits^{i\Phi_{\pm}t}+\mathrm{h.c.}\right)  \mathrm{{,}} \label{3}%
\end{equation}
where the phase $\Phi_{\pm}=\pm\left\vert \Omega_{1}\right\vert \left(
\frac{\left\vert \widetilde{\lambda}_{b}\right\vert ^{2}}{4\left\vert
\Omega_{1}\right\vert ^{2}-\delta_{b}^{2}}-\frac{\left\vert \widetilde
{\lambda}_{a}\right\vert ^{2}}{4\left\vert \Omega_{1}\right\vert ^{2}%
-\delta_{a}^{2}}\right)  +\delta_{a}-\delta_{b}$ can be made null only in the
intermediate regime where the term in brackets, multiplied by $\left\vert
\Omega_{1}\right\vert $, can be made of the order of the detuning $\delta
_{a}-\delta_{b}$. The coupling parameters $\Sigma_{\pm W}=\widetilde{\lambda
}_{a}\widetilde{\lambda}_{b}^{\ast}\left(  \delta_{a}-\delta_{b}\right)
/2\delta_{a}\delta_{b}$, $\Sigma_{\pm I}=\pm\widetilde{\lambda}_{a}%
\widetilde{\lambda}_{b}^{\ast}\left\vert \Omega_{1}\right\vert /\left(
4\left\vert \Omega_{1}\right\vert ^{2}-\delta_{a}\delta_{b}\right)  $, and
$\Sigma_{\pm S}=\pm\widetilde{\lambda}_{a}\widetilde{\lambda}_{b}^{\ast
}/4\left\vert \Omega_{1}\right\vert $, follows without any need to adjust the
detuning $\delta_{1}$, since the condition $\delta_{1}\ll\left\vert \Omega
_{1}\right\vert $,$\left\vert \delta_{a}\right\vert $,$\left\vert \delta
_{b}\right\vert $ is satisfied. Again, the strength of the coupling parameters
obey the relation $\Sigma_{I}\gtrsim\Sigma_{W}\gtrsim\Sigma_{S}$. The
time-dependent Hamiltonian (\ref{3}) can be treated through the invariants
introduced by Lewis and Riesenfeld \cite{LR}, as discussed in Ref.
\cite{CFRM,CNRM}. Otherwise, we may consider identical modes $\omega
_{a}=\omega_{b}$ (so that $\delta_{a}=\delta_{b}$) of two identical cavities
disposed along perpendicular transversal axes and sharing the same two-level atom.

\subsection{\textit{Applications}}

As mentioned above, the engineered bilinear Hamiltonians can be used for
quantum state preparation in cavity QED. In Refs. \cite{Roberto,CNRM} the
interaction (\ref{2}) was employed in a protocol for the preparation of the
original Einstein-Podolsky-Rosen (EPR) entanglement expanded in the Fock
representation. This interaction was also required to engineer the even and
odd EPR states defined in Ref. \cite{Celso2}.{\ The advantage of the protocol
in \cite{Celso2} over that in \cite{Roberto} is the use of a intense classical
amplification field, where the strength of the bilinear interactions between
the cavity modes are considerably increased, at least by one order of
magnitude, compared to the strength in \cite{Roberto}}. Consequently, the
atom-field interaction time required to obtain high-fidelity states can be
considerably shorter,\ making the dissipative effects negligible.

The present scheme, in turn, has the advantage over those in Refs.
\cite{Roberto,Celso2}, in that a two-level atom is able to generate both
interactions, (\ref{2}) and (\ref{3}), with coupling strengths comparable to
those obtained in \cite{Celso2}. Therefore, apart from the benefit of a
shorter interaction time, due to the strength of the coupling parameters
$\Lambda$ and $\Sigma$, here we get an additional advantage employing circular
long-lived Rydberg states. The same facilities apply to the squeezing
Hamiltonian engineered below. We finally mention that, following the reasoning
in Ref. \cite{Pittman}, our engineered bilinear interactions can be considered
to manipulate quantum information in cavity QED.

\section{\textit{The Squeezing Hamiltonian }}

To obtain the parametric amplification Hamiltonian, we consider, as indicated
in Fig. 2, the atomic transition coupled to a single cavity mode ($\omega_{a}%
$), as well as two classical amplification fields, with $\delta_{1}=0$ and
$\delta_{2}<0$, under the condition $\left\vert \Omega_{1}\right\vert
=-\delta_{2}/2\gg\left\vert \widetilde{\lambda}_{a}\right\vert ,\left\vert
\Omega_{2}\right\vert ,\left\vert \delta_{a}\right\vert $. Starting from the
interaction picture we obtain, after the unitary transformation $U_{1}%
=\exp\left[  -i\left(  \Omega_{1}\sigma_{eg}+\Omega_{1}^{\ast}\sigma
_{ge}\right)  t\right]  $ and within the rotating wave approximation, the
interaction%
\begin{equation}
\widetilde{V}_{1}(t)=\widetilde{\lambda}_{a}a\operatorname*{e}%
\nolimits^{i\delta_{a}t}\left(  \sigma_{++}-\sigma_{--}\right)  /2-\widetilde
{\Omega}_{2}\sigma_{+-}+\mathrm{h{.c.}}, \label{3a}%
\end{equation}
where $\widetilde{\Omega}_{2}=\Omega_{2}\operatorname*{e}\nolimits^{-i\varphi
_{1}}/2\mathrm{{.}}$ Through the new basis $\left\{
\genfrac{\vert}{\rangle}{0pt}{}{\uparrow}{\downarrow}%
=\left(  \operatorname*{e}\nolimits^{i(\varphi_{2}-\varphi_{1})}\left\vert
+\right\rangle \pm\left\vert -\right\rangle \right)  /\sqrt{2}\right\}  $ it
is straightforward to verify that, after another transformation $U_{2}%
=\exp\left[  i\left(  \widetilde{\Omega}_{2}\sigma_{+-}+\mathrm{h{.c.}%
}\right)  t\right]  $, the interaction (\ref{3a}) becomes%
\begin{equation}
\widetilde{V}_{2}(t)=\left(  \widetilde{\lambda}_{a}a\operatorname*{e}%
\nolimits^{i\delta_{a}t}+\widetilde{\lambda}_{a}^{\ast}a^{\dagger
}\operatorname*{e}\nolimits^{-i\delta_{a}t}\right)  \left(  \sigma
_{\uparrow\downarrow}\operatorname*{e}\nolimits^{-i\left\vert \Omega
_{2}\right\vert t}+\sigma_{\downarrow\uparrow}\operatorname*{e}%
\nolimits^{i\left\vert \Omega_{2}\right\vert t}\right)  , \label{3b}%
\end{equation}
which is suitable for the derivation, under the assumption that $\left\vert
\delta_{a}\right\vert \ll\left\vert \Omega_{2}\right\vert $, of the effective
Hamiltonian $\mathcal{H}=-i\widetilde{V}_{2}(t)\int\widetilde{V}_{2}%
(t^{\prime})dt^{\prime}$. Finally, for the initial atomic state $\left\vert
\genfrac{.}{.}{0pt}{}{\uparrow}{\downarrow}%
\right\rangle $ and adjusting $\delta_{a\binom{\uparrow}{\downarrow}}%
=\mp2\left\vert \widetilde{\lambda}_{a}\right\vert ^{2}/\left\vert \Omega
_{2}\right\vert $, we obtain the engineered parametric amplification Hamiltonian%

\begin{equation}
\mathcal{H}_{_{\binom{\uparrow}{\downarrow}}}=\mp\chi\left(  \operatorname*{e}%
\nolimits^{-2i\varphi_{1}}a^{2}+\mathrm{h.c.}\right)  \mathrm{{,}} \label{4}%
\end{equation}
where $\chi=\left\vert \widetilde{\lambda}_{\alpha}^{2}/4\Omega_{2}\right\vert
$, which allows the squeezing of any desired prepared cavity-field state. We
note that the squeezing direction in phase space is controlled through the
phase factor $\operatorname*{e}\nolimits^{-i\varphi_{1}}$ derived from a
classical amplification field. With $\varphi_{1}=\varphi_{2}=0$ we recover the
atomic bases $\left\{  \left\vert g\right\rangle ,\left\vert e\right\rangle
\right\}  $ in that $\left\vert \uparrow\right\rangle =\left\vert
e\right\rangle $ and $\left\vert \downarrow\right\rangle =\left\vert
g\right\rangle $. Differently from the protocols with three-level atoms
\cite{CNRM,Roberto,Celso2,Zagury}, where the squeezing interaction is
engineered through degenerate atomic transitions, here a second amplification
field on a single atomic transition is required to achieve the two-photon process.

\subsection{\textit{Applications}}

In the running-wave domain, the squeezed states revealed the intrinsic quantum
nature of light, together with direct evidence for an atom undergoing a
quantum jump \cite{Milburn}. The engineered interaction (\ref{4}) exposes a
myriad of possible applications in cavity QED, ranging from the preparation of
a set of squeezed states \cite{NRCM} to the possibility of revealing the
statistical properties of the electromagnetic field through its controlled
interaction with atoms. Beyond these applications, a particular squeezed
superposition state (SSS) can be prepared when adjusting $\delta_{a}%
=\varphi_{1}=\varphi_{2}=0$, such that the Hamiltonian governing the evolution
of the atom-field state reads%
\begin{equation}
\mathcal{H}=-\chi\left[  2a^{\dagger}a+a^{2}+\left(  a^{\dagger}\right)
^{2}\right]  \left(  \sigma_{ee}-\sigma_{gg}\right)  \mathrm{{.}} \label{5}%
\end{equation}
Starting from the initial state $\left(  \left|  e\right\rangle +\left|
g\right\rangle \right)  \left|  \alpha\right\rangle /\sqrt{2}$, $\left|
\alpha\right\rangle $\ being a coherent state injected into the cavity, the
generated SSS is $\left(  \left|  e\right\rangle U_{e}+\left|  g\right\rangle
U_{g}\right)  \left|  \alpha\right\rangle /\sqrt{2}$, where $U_{\ell}$\ stands
for the evolution operator associated with Hamiltonian $H_{\ell}=\left\langle
\ell\right|  \mathcal{H}\left|  \ell\right\rangle $, with $\ell=e,g$.\ It has
been shown in Ref.\cite{CFRM} that the decoherence time of this particular SSS
-- where both states composing the superposition exhibt the same squeezing
direction -- could be delayed to around the relaxation time of the cavity
field. This remarkable result requires the engineering of absolute zero
reservoirs composed by oscillators squeezed in a direction perpendicular to
that of the superposition state. The reason behind this phenomenon is quite
palpable: the injection of noise from the reservoir into the superposition
state decreases as the degree of squeezing, of both the reservoir and the
superposition states, increases. Therefore, the present scheme is a crucial
step towards the accomplishment of this specific program for protecting a
quantum state.\textbf{\ }(We stress that the engineering of an ideally
squeezed reservoir for a cavity mode, a task only partially achieved in Ref.
\cite{Vitali}, has also been accomplished by our group \cite{FMV}.)

\section{\textit{The AJC Hamiltonian}}

To engineer the AJC model in cavity QED we start from the interaction
$\widetilde{V}_{2}(t)$ that leads to the squeezing operator. Considering
$\varphi_{1}=\varphi_{2}=0$ and $\delta_{a}=-\left\vert \Omega_{2}\right\vert
$, we obtain, within the rotating-wave approximation following from the
condition $\left\vert \Omega_{2}\right\vert \gg\left\vert \widetilde{\lambda
}_{a}\right\vert ^{2}$, the desired interaction%
\begin{equation}
\mathcal{H}_{AJC}=\widetilde{\lambda}_{a}a\sigma_{ge}+\mathrm{h.c.}%
\mathrm{{,}} \label{6}%
\end{equation}
which has already been engineered in cavity QED with two-level \cite{Solano}
or three-level atoms \cite{Ruynet}.{\large \ }While in Ref. \cite{Solano} the
AJC interaction is achieved, with a single classical field, by adjusting the
detuning between the cavity mode and the atomic transition, in the present
scheme we adjust the detuning between the classical field $\omega_{2}$ and the
atomic transition. It is also possible to obtain an alternation between the JC
and the AJC model assuming the same parameters that lead to the AJC model
($\varphi_{1}=\varphi_{2}=0$, $\left\vert \delta_{a}\right\vert =-\left\vert
\Omega_{2}\right\vert $, and $\left\vert \Omega_{2}\right\vert \gg\left\vert
\widetilde{\lambda}_{a}\right\vert $). To this end, two synchronized pulsed
fields must be introduced, leading to $\omega_{2}(t)=\omega_{0}-$ $\delta
_{2}(t)$, with $\delta_{2}(t)=\left\vert \delta_{2}\right\vert \left[
\Theta_{1}(t)-\Theta_{2}(t)\right]  $. Given $\Theta_{\ell}(t)=\sum_{n}%
\Theta\left[  t-\left(  2n+\delta_{\ell2}\right)  \tau\right]  \Theta\left[
(2n+1+\delta_{\ell2})\tau-t\right]  $, $\tau$ being the duration of each pulse
and $n$ an integer, the engineered Hamiltonian reads%
\begin{equation}
\mathcal{H}=\Theta_{1}(t)\left(  \widetilde{\lambda}_{a}a\sigma_{ge}%
+\mathrm{h.c.}\right)  +\Theta_{2}(t)\left(  \widetilde{\lambda}_{a}%
a\sigma_{eg}+\mathrm{h.c.}\right)  \mathrm{{.}} \label{7}%
\end{equation}

\section{Discussion and Conclusion}

Next, addressing some sensitive points in the present engineering scheme, we
turn our attention to the effective squeezing interaction in Eq. (\ref{4}), to
demonstrate that it follows with good agreement from the full Hamiltonian in
Eq. (\ref{1}). In Fig. 3, starting with the cavity field in the vacuum state
and the atom in the ground state, we plot the variance of the cavity-field
squeezed quadrature, $\left(  \Delta X\right)  ^{2}$, against the squeezing
factor $r=2\chi t$. Although the present scheme can be applied to both optical
and microwave regimes, in Fig. 3 we have used typical parameters from cavity
QED experiments in the\ microwave regime, which, in units of $\lambda
_{a}=3\times10^{5}$s$^{-1}$, are approximately given by: $\omega_{0}=10^{5}$,
$\Omega_{1}=4\times10^{2}$, $\Omega_{2}=20$, $\delta_{2}=8\times10^{2}$. The
solid line corresponds to the variance computed either analytically or
numerically from the effective squeezing interaction (\ref{4}) (the
exceedingly small difference between the two curves being around $0.3\%$ for
$r=1$). The dashed line corresponds to the variance computed numerically from
the full Hamiltonian in Eq. (\ref{1}). Fig. 3 reveals a good agreement between
the solid and dashed lines, for $r$ ranging from zero to unity, with the
degree of squeezing being $86.4\%$ and $85.6\%$, respectively. Evidently,
these degrees of squeezing can even be enhanced considering a sample of $N$
noninteracting atoms instead of a single one, as considered in \cite{Zagury}.

We also computed the effects of \ both dissipative mechanisms, the cavity
field and the atom, on the degree of squeezing achieved. The dotted line
traces the numerical computation of the variance of the squeezed quadrature
based on the master equation
\begin{equation}
\overset{.}{\rho}=-i\left[  H,\rho\right]  +\mathcal{L}_{field}\rho
+\mathcal{L}_{atom}\rho\mathrm{{,}} \label{8}%
\end{equation}
with $H$ given by Eq. (\ref{1}). As usual for a reservoir at absolute zero,
the Liouville operators read: $\mathcal{L}_{field}\bullet=\left(  \Gamma
_{f}/2\right)  \left(  2a\bullet a^{\dagger}-a^{\dagger}a\bullet-\bullet
a^{\dagger}a\right)  $ and $\mathcal{L}_{atom}\bullet=\left(  \Gamma
_{a}/2\right)  \left(  2\sigma_{-}\bullet\sigma_{+}-\sigma_{+}\sigma
_{-}\bullet-\bullet\sigma_{+}\sigma_{-}\right)  $. We again assumed, in units
of $\lambda_{a}$, typical values for high-finesse cavities and
circular-Rydberg states, around $\Gamma_{f}=3\times10^{-3}$ and $\Gamma
_{a}=10^{-4}$. From Fig. 3 we observe that the degree of squeezing under the
dissipative effects, falls to $80.5\%$, which is still a remarkable result.

In this work we have presented protocols to build bilinear and quadratic
Hamiltonians in cavity QED, employing a single two-level atom plus classical
amplification fields. The simplicity and generality of these schemes make them
suitable for the implementation of quantum logical operations \cite{Pittman},
quantum state preparation \cite{Celso2}, and fundamental tests of quantum
theory \cite{Rausch}.\textbf{\ }The validity of the approximations leading to
our effective Hamiltonians has been confirmed with numerical calculations,
even under the effect of dissipative mechanisms. As well as deepening our
understanding of atom-field interaction in cavity QED, our protocols may also
open the way to advances in correlated areas such as trapped-ion and
nanomechanical oscillators.

\textbf{Acknowledgments}

We wish to express thanks for the support from FAPESP (Projeto Tem\'{a}tico
N\'{u}mero 00/15084-5), CNPq (Instituto do Mil\^{e}nio de Informa\c{c}\~{a}o
Qu\^{a}ntica), and CAPES, Brazilian agencies, and to Tim Roberts for a
critical reading of the manuscript.

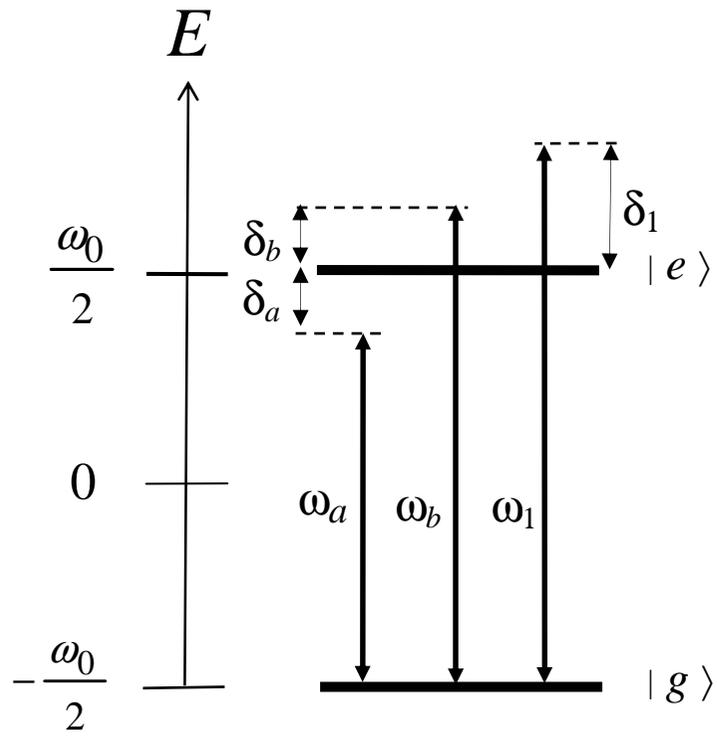
Fig. 1. Energy diagrams underlying the bilinear Hamiltonians (a)
$ab+\mathrm{h.c.}$ and (b) $ab^{\dagger}+\mathrm{h.c.}$

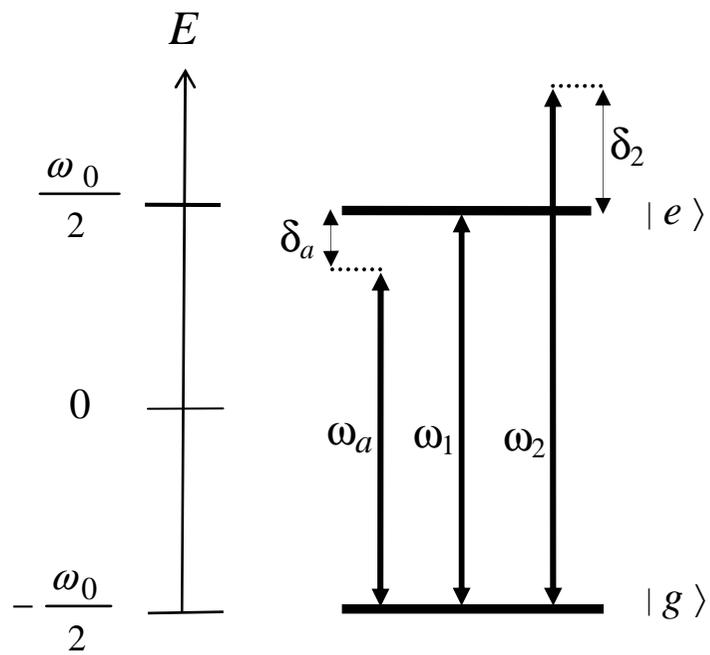
Fig. 2. Energy diagram of the scheme used to obtain the squeezing interaction.

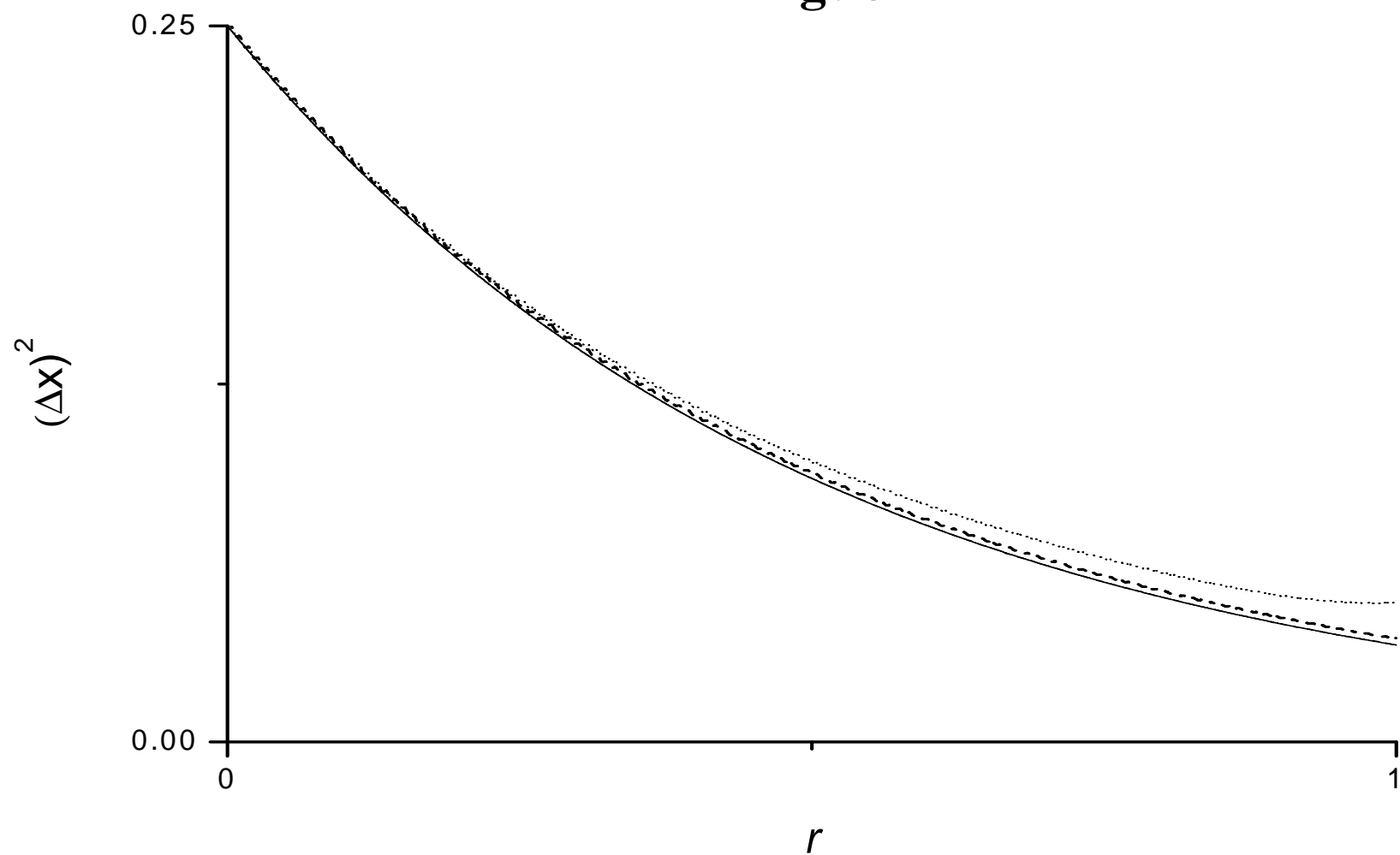
Fig. 3. Variance of the cavity-field squeezed quadrature $\left(  \Delta
X\right)  ^{2}$ plotted against the squeezing factor $r$ for the effective
squeezing interaction in Eq. (\ref{4}) (solid line), the full Hamiltonian in
Eq. (\ref{1}) (dashed line), and the full Hamiltonian under cavity-field and
atomic decay (dotted line).
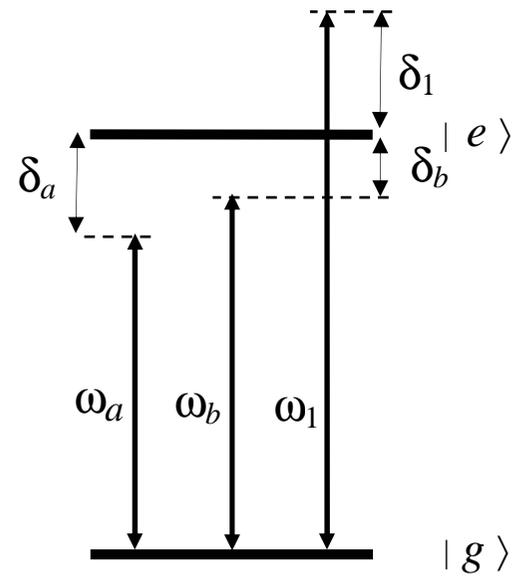
\end{document}